\documentclass{article}
\usepackage{amsfonts,amsmath,amsxtra,graphicx}
\usepackage{mathtext}
\usepackage{amsmath,amsfonts,amsxtra, mathtools}
\usepackage{amssymb}
\usepackage{amscd}
\usepackage[matrix,arrow,curve]{xy}

\newcounter{N}
\newcommand\eqq[1]{\begin{eqnarray}
\begin{split}
#1
\end{split}
\end{eqnarray}}

\def\be{\begin{eqnarray}}
\def\ee{\end{eqnarray}}
\def\nn{\nonumber}

\def\p{\partial}
\def\lm{\limits}

\def\op{\hat \nabla}
\def\opc{\check \nabla}

\hoffset= - 1.0in         % switch off for draft style
\title{{\bf $S$-Duality and Modular Transformation as
a non-perturbative deformation of
the ordinary $pq$-duality} \vspace{.2cm}}
\author{{\bf D.Galakhov}\thanks{{\small
{\it ITEP, Moscow, Russia} and {\it NHETC and Department of Physics and Astronomy, Rutgers University,
Piscataway, NJ 08855-0849, USA }};
galakhov@itep.ru, galakhov@physics.rutgers.edu }, {\bf A.Mironov}\footnote{ {\small {\it Lebedev
Physics Institute} and {\it ITEP, Moscow, Russia}}; mironov@itep.ru;
mironov@lpi.ru}, {\bf A.Morozov}\thanks{{\small {\it ITEP, Moscow,
Russia}}; morozov@itep.ru}
\date{ }}
\date{\today}

\begin{document}
\maketitle
\vspace{-5.0cm}

\begin{center}
\hfill FIAN/TD-15/13\\
\hfill ITEP/TH-42/13\\
\end{center}

\vspace{3.5cm}

\begin{abstract}
A recent claim that the $S$-duality between $4d$
SUSY gauge theories, which is AGT related to the modular
transformations of $2d$ conformal blocks,
is no more than an ordinary Fourier transform
at the perturbative level,
is further traced down to the
commutation relation $[\check P, \check Q]=-i\hbar$
between the check-operator monodromies of the exponential
resolvent operator in the underlying Dotsenko-Fateev
matrix models and $\beta$-ensembles.
To this end, we treat the conformal blocks as
eigenfunctions of the monodromy check operators, what is
especially simple in the case of one-point toric block.
The kernel of the modular transformation is then
defined as the intertwiner of the two monodromies,
and can be obtained straightforwardly, even when the eigenfunction
interpretation of the blocks themselves is technically tedious.
In this way, we provide an elementary derivation of
the old expression
for the modular kernel for the one-point toric
conformal block.
\end{abstract}

%\tableofcontents

\section{Introduction}

$S$-duality is one of the most interesting discoveries
of modern string theory \cite{Sdu}.
It is a far-going generalization of the
$E-B$ duality of Maxwell electrodynamics with magnetic charges
which states that the non-perturbative partition functions of
different field theories can coincide after a non-linear
transformation of coupling constants.
$S$-dual theories can have different numbers of
perturbative degrees of freedom, different gauge groups
and, perhaps, even live in different space-time dimensions.
A significant class of $S$-dual models can be described
by the $M5$-brane construction of \cite{Mbrane,Mikh,Ga},
where $6d$ theory on the brane is compactified on a
$2d$ Riemann surface, which therefore controls
the structure of emerging perturbative $4d$ gauge theory,
and thus provides a natural explanation of the
hidden integrable structure \cite{GKMMM,ITEP},
the Riemann surface being just the spectral surface of an integrable system.
In this picture $S$-dualities get related to modular
transformations of the Riemann surface.
A quantitative realization of this idea \cite{Ga}
led to the AGT conjecture \cite{AGT,AGT1},
which identifies the LMNS instanton sums \cite{LMNS},
expressed via Nekrasov functions \cite{Nek},
with the conformal blocks of $2d$ conformal theories \cite{CFT}.
This identification opens a way to a quantitative study of
$S$-dualities, because constructing modular transformations of conformal
blocks is a hard but still solvable problem.

The problem is that the original definition provides
conformal blocks in a form of perturbative series in
the variable $x$; in the case of the spherical $4$-point
conformal block, AGT related to the ordinary $SU(2)$ SYM theory
with $4$ hypermultiples, $x$ is just a double ratio of the
four punctures, and the modular transformation relates
the conformal blocks at points $x$ and $1-x$.
Normally one needs some non-perturbative completion of
this definition to even pose the problem.

There is a variety of such definitions: exploiting an $SL_q(2)$ counterpart of
conformal blocks in specific representations \cite{PT,Tesch}
or various equations that they can satisfy, from
Ward identities for extended blocks with additional
insertions of degenerate fields \cite{AGGTV,DGOT,Wy,Maru,Kanno,MMM}
to wonderful, still badly understood, relations to
the Painlev\'e IV equation \cite{ILT}.

Remarkably, study of the AGT relations provides as a byproduct
a kind of a more direct approach.
The conformal blocks possess a matrix model ($\beta$-ensemble)
realization \cite{AGTmamo,three-point,AGTmamoMM},
which is an advanced version
of the old Dotsenko-Fateev trick \cite{DF}
and the Felder construction \cite{Feld}
(in particular, the integration contours for
screening charges are actually open, not always closed,
but instead one suffices to use only one screening charge of two).
Then one can study the genus expansion in this theory,
which is actually preserved by $S$-duality.
This can be also considered as studying $S$-duality
for expansions at the point $g_s=0$,
where all terms are explicit "non-perturbative"
functions in $x$.

In fact, one should be  careful with the word "perturbative"
in the present context.
In the above mentioned standard definition, the conformal block
is a perturbative series not only in $x$, but also in the
dimensions of operators,
i.e. in the string coupling constant
$g_s^2 = \epsilon_1\epsilon_2$.
Non-perturbative corrections exist both in $x$ and in $g_s$,
and they are actually very different.
In what follows we reserve the words "perturbative"
and "non-perturbative" for  the $g_s^2$-corrections,
assuming that the $x$-behavior is completely fixed by
switching to the matrix model description.
Thus, "the perturbative conformal block" refers to the genus
expansion of the  $\beta$-ensemble controlled by the
"topological recursion" formalism of \cite{AMM}-\cite{AMMlast}
and \cite{CheE},
while "the non-perturbative conformal block" refers to a  more obscure
quantity, which still does not have a unique commonly
accepted definition.
Hopefully, all the existing suggestions, \cite{PT,ILT,AGGTV,DGOT}
and the one described in the present paper,
would finally lead to the same outcome,
but this still remains to be demonstrated and understood.

\bigskip

Anyhow, in what follows we concentrate on one particular definition,
that of the Dotsenko-Fateev $\beta$-ensemble of \cite{AGTmamoMM}
and use it to study the $S$-dualities (modular transformations)
of both perturbative and non-perturbative conformal blocks.

Making use of this idea and calculational advances
in AGT studies, we conjectured recently \cite{GMM,Nem} that {\bf the $S$-duality
is actually reduced to an ordinary Fourier transform in all orders
of perturbation expansion in string coupling constant}.
In \cite{GMM} this was shown for the central charge
$c= 1 - 6(\beta-1/\beta)^2=1$ ($\beta=1$),
and, after a more accurate analysis of normalization
factors in \cite{Nem}, this result was extended to
an arbitrary $\beta$.
This claim was reconsidered and confirmed from a
slightly different viewpoint in \cite{it1,it2}.
These results are perturbative,
and their exact relation to non-perturbative
suggestions of \cite{PT,ILT} still remained obscure,
despite the latter formulas are also consistent with
the pure Fourier transform at the perturbative level.

\bigskip

Calculations of \cite{GMM,Nem} are quite tedious,
what seems strange for such a simple outcome.
Clearly some very simple explanation should exist,
which does not require long calculations.
It is the purpose of the present paper to provide such
an explanation in precise and quantitative form. This calls for
begins an investigation of the far-reaching
corollaries of emerging formalism, which so
far seemed to be just a funny technical tool
in advanced matrix model theory \cite{AMMcheck}.

An intuitive idea has already been formulated in \cite{AGGTV}:
to describe the duality, one can treat dual conformal blocks
as eigenfunctions of canonically conjugated quantum operators.
The question is what are the operators and how they
act on the correlation functions,
and it is where the matrix model theory is of a great use.
Namely, it puts the story into the context of Seiberg-Witten (SW) theory,
where the partition function is defined as a function of flat moduli $\vec a$ by the equations
\be
\left(\oint_{\vec A} \lambda\right)  Z(\vec a) = \vec a Z(\vec a), \nn \\
\left(\oint_{\vec B} \lambda\right) Z(\vec a)= \frac{\partial }{\partial \vec a}Z(\vec a)
\ee
which allows one to treat the periods of the SW differential as operators acting on functions on the moduli space.
In the genus expansion of matrix models the role of SW differential is played by the one-point
resolvent, which can be defined either in the usual style of \cite{AMM} or alternatively reformulated 
as produced by the so-called check-operators \cite{AMMcheck} which act on ramification points of the spectral curve. 
Accordingly,
\begin{list}{\roman{N})}{\usecounter{N}}
\item the period integrals of the resolvent generating operators turn out
to establish a set of canonically conjugated observables \cite{AMMcheck},
and
\item partition functions are their eigenfunctions.
\end{list}
In fact, there are delicate points in this story.
The genus expansion is the typical quasiclassical expansion,
thus, it actually suffers from the Stokes like phenomena,
which requires a careful interplay between different branches
of the Seiberg-Witten differentials.
Taking this into account provides a natural non-perturbative
completion of the genus expansion, and can be used as yet
another definition of the non-perturbative conformal blocks
and non-perturbative modular transformations.
We demonstrate that in the simplest examples the results
seem consistent with the ansatz of \cite{PT,Tesch}.
An additional advantage of such an approach is a clear relation to
the theory of wall crossing {\it a la} \cite{GMN},
to the cluster algebras \cite{FG}  and
to the Kontsevich-Soibelman formulas \cite{KS}.
We elaborate more on these relations in a separate text.

\section{Duality and eigenfunctions of dual operators}
\label{sec:Duality}

An archetypical example of duality is provided by the
switch between coordinate and momentum operators.
Namely, consider the two operators
$\hat A = e^{i\hat {\cal P}}$ and $\hat B = e^{i\hat {\cal Q}}$,
with the commutation relation
\be
\hat A\hat B = e^{i\hbar}\hat B\hat A
\label{ABPQ}
\ee
Then, their eigenfunctions are related by the Fourier transform
in the eigenvalue space:
\be\begin{array}{c}
\hat A Z_a({\cal Q}) = e^{ia}Z_a({\cal Q})\\  \\
\hat B \tilde Z_{a'}({\cal Q})=e^{ia'}\tilde Z_{a'}({\cal Q})
\end{array} \ \ \ \ \ \ \ \ \ \ \ \ \ \
\stackrel{(\ref{ABPQ})}{\Longrightarrow} \ \ \ \ \ \ \
Z_a({\cal Q}) = \int e^{\frac{iaa'}{\hbar}}\tilde Z_{a'}({\cal Q})da'
\ee
This can be easily checked in this case by calculating the
eigenfunctions explicitly:
\be
Z_a({\cal Q}) = e^{\frac{ia {\cal Q}}{\hbar}}, \ \ \ \ \ \ \
\tilde Z_{a'}({\cal Q}) = \delta({\cal Q}-a')
\ee
but this is {\bf not necessary} in order to define what is the transformation kernel.
Instead, one can substitute the two operators by their representatives
in the eigenvalue space, which reproduce the right commutation relations:
\be\label{check}
\check A = e^{ia}, \ \ \ \ \ \
\check B = e^{{\hbar}\frac{\partial}{\partial a}}
\ee
which we call {\it check-operators}, following \cite{AMMcheck}.
Then the transformation kernel $M(a,a')=e^{\frac{iaa'}{\hbar}}$
is simply defined from the
relation
\be\label{kernel}
\boxed{
\check A(a) M(a,a') = \check B(a')M(a,a')
}
\ee
Here a delicate point is the possibility to multiply the operator $B$ in (\ref{check}) by an arbitrary
function of $a$: this normalization factor requires attention in more
sophisticated examples below.

This is the approach to duality transformations, which we are going to apply
in general. That is, we will construct a pair of peculiar operators $\hat A$ and $\hat B$ such that
the conformal block, i.e. the matrix model ($\beta$-ensemble) partition function
is an eigenfunction of $\hat A$, while the modular transformed conformal block is an eigenfunction of $\hat B$.
Then, the AGT correspondence guarantees the same relation between the $S$-dual $N=2$ supersymmetric gauge theories.

As we shall see, the only difference of the modular $S$-duality from the above
ordinary $pq$-duality is that the relevant operators $\hat A$ and $\hat B$
in the case of
{\it non-perturbative} conformal blocks commute in a little less trivial
way, but {\it perturbatively} they satisfy exactly (\ref{ABPQ}).
This explains the perturbative result of \cite{GMM,Nem}
for the {\it properly normalized} conformal blocks
and straightforwardly provides its non-perturbative generalization,
which is in accordance with \cite{PT}.

\section{Modular transformations: conformal blocks and $\beta$-ensembles}
\label{sec:Blocks}

The infinite conformal symmetry in two dimensions allows one to expand any correlator in CFT into the conformal blocks \cite{CFT}.
We consider here the one-point correlator on torus. The corresponding conformal block is usually
represented by a series in the torus modular parameter $q=e^{\pi i \tau}$ with coefficients depending on the external dimension
$\Delta_{\rm ext}$ and on the intermediate dimension $\Delta$. Hereafter, we use the following useful parametrization of the
CFT quantities
\be
\Delta=\frac{Q^2}{4}-a^2,\quad \Delta_{\rm ext}=\mu(Q-\mu),\quad c=1+6Q^2, \quad Q=b+b^{-1}
\ee
We assume the conformal block to be normalized as follows
\be
B_a(\tau|\mu)=1+q\left(\frac{\Delta_{\rm ext}(1-\Delta_{\rm ext})}{2\Delta}+1\right)+O(q^2)
\ee
Throughout the paper we rescale the conformal dimensions to include $g_s^2$ so that
the perturbative series correspond to the large $a$ expansions.

The $6j$-symbols (the Racah coefficients) for the Virasoro algebra can be realized as the fusion relation connecting
the conformal blocks at modular transformed moduli of the torus:
\be
B_a(\tau|\mu)=\int db M(a,b) B_b\left(-\frac{1}{\tau}\Big|\mu\right)
\ee

The conformal block can be related to the elliptic $\beta$-ensemble partition function \cite{MMtorus}
\be\label{be}
Z_a(\tau|\mu)=q^{-a^2}\int\lm_0^{\pi}dz_1\ldots\int\lm_0^{\pi}dz_N\; \prod\lm_{i<j}\theta(z_i-z_j)^{-2b^2}\prod\lm_i \theta(z_i)^{-2b\mu}
 e^{-4 i a\left(\sum\lm_i b z_i +\mu w\right)},
\ee
with the number of integrals constrained by the condition $\mu+bN=0$.
Here the toric heat kernel reads
\be\label{eq:Green}
\theta(z)=2 q^{\frac{1}{8}}\sin z\; \prod\lm_{n=1}^{\infty}(1-q^n)(1-2q^n\cos 2z+q^{2n})=\sum\lm_{n=0}^{\infty}(-1)^n q^{n(n+1)/2}\sin(2n+1)z
\ee
The concrete relation of this partition function and the conformal block is described by the formula\footnote{We choose the
Dedekind function to be $\eta(q)=q^{\frac{1}{24}}\prod\lm_n(1-q^n)$.}
\be
Z_a(\tau|\mu)=\frac{Z_a(i\infty|\mu)}{\eta(q)^{\nu}}B_a(\tau|\mu),\quad \nu=3\Delta_{\rm ext}+3N-1
\ee
The claim of \cite{GMM,Nem} was that for any set of parameters %(which we denote $\mathfrak{P}$)
\be
Z_a(\tau|\mu)=\int d b e^{2\pi i a b}Z_b\left(-\tau^{-1}|\mu\right)
\ee
at any perturbative order in $\mu/a$. Here the conformal block $Z_a(\tau|\mu)$  and its modular transformed $Z_b\left(-\tau^{-1}|\mu\right)$ play the role of $Z_a({\cal Q})$ and $\tilde Z_b({\cal Q})$ of s.2 correspondingly.

\section{Modular transformation of $\beta$-ensemble: perturbative level}
\subsection{Loop equations and their symmetries}
\label{sec:loop}

The key role in our consideration is played by the resolvent operator. A net definition for the $n$-point resolvent
for the $\beta$-ensemble on some generic Riemann surface can be given as an average
\be
R_n(\xi_1,\ldots,\xi_n)=\left\langle\left(\sum\lm_{i_1}\frac{E'(\xi_1,z_{i_1})}{E(\xi_1,z_{i_1})}\right)\ldots\left(\sum\lm_{i_n}\frac{E'(\xi_n,z_{i_n})}{E(\xi_n,z_{i_n})}\right)\right\rangle
\ee
over a $\beta$-ensemble like (\ref{be}), with $x_i$ being integration variables in the $\beta$-ensemble and the prime means differentiating with respect to the first argument.
Here $E(z,w)$ is the prime form \cite{Fay}, its logarithm plays a role
of Green function for the scalars;
in our particular toric case it is given by expression (\ref{eq:Green}) (up to inessential constant which cancels out in the ratio).

One can introduce (infinitely many) additional time variables, in order to generate
the multi-point disconnected resolvents by an operator acting on these times, \cite{AMM} so that
\be
R_n(\xi_1,\ldots,\xi_n)=Z^{-1}\op(\xi_1)\ldots \op(\xi_n) Z
\ee
Similarly, one introduces a set of connected resolvents
\be
\rho_n(\xi_1,\ldots,\xi_n)=\op(\xi_1)\ldots \op(\xi_n) \log Z
\ee

Following \cite{toric_loop} an infinite set of Ward identities for the $\beta$-ensemble partition function can be derived
in a simple way by the shift of the integration variables
\be
z_i\rightarrow z_i+\epsilon \p_\xi \log \theta(\xi-z_i)
\ee

Thus at the first $\epsilon$-order one derives an identity
\eqq{
\left\langle\left(\sum\lm_i \frac{\theta'(\xi-z_i)}{\theta(\xi-z_i)}\right)^2-
\sum\lm_i \frac{\theta''(\xi-z_i)}{\theta(\xi-z_i)}+(-2b^2\mu\p_{\xi}\log\theta(\xi-w)+
4 i b^2 a)\sum\lm_i \frac{\theta'(\xi-z_i)}{\theta(\xi-z_i)}+\right.\\
\left.+2\mu b \sum\lm_i \frac{\theta'(\xi-z_i)}{\theta(\xi-z_i)}
\left(\sum\lm_i \frac{\theta'(z_i-w)}{\theta(z_i-w)}-\sum\lm_i \frac{\theta'(\xi-w)}{\theta(\xi-w)}\right)-
2b^2\sum\lm_{i<j} \frac{\theta'(z_i-z_j)}{\theta(z_i-z_j)}\left(\sum\lm_i \frac{\theta'(\xi-z_i)}{\theta(\xi-z_i)}-
\sum\lm_i \frac{\theta'(\xi-z_j)}{\theta(\xi-z_j)}\right)\right\rangle=0\nn
}
Using the relation
\be
\begin{split}
\frac{\theta'(x-y)\theta'(x-z)}{\theta(x-y)\theta(x-z)}+\frac{\theta'(y-x)\theta'(y-z)}{\theta(y-x)\theta(y-z)}+\frac{\theta'(z-x)\theta'(z-y)}{\theta(z-x)\theta(z-y)}=\frac{1}{2}\left(\frac{\theta''(x-y)}{\theta(x-y)}+
\frac{\theta''(y-z)}{\theta(y-z)}+\frac{\theta''(y-z)}{\theta(y-z)}\right)+3\eta_1\nn
\end{split}
\ee
where $\eta_1=4\frac{\p\log\eta}{\p \log q}$, and after a little algebra one can derive the following loop equation
\eqq{
-b^2\langle R(\xi,\xi)\rangle -Qb\langle R'(\xi)\rangle+(-2\mu \p_\xi\log\theta(z-w)-4 i a)b\langle R(\xi)\rangle+3b\mu(N+1)\eta_1-\\
-\frac{\theta'(\xi-w)}{\theta(\xi-w)}\p_w\log Z-bN\mu\frac{\theta''(\xi-w)}{\theta(\xi-w)}+4\frac{\p \log Z}{\p \log q}+4a^2-4ia \mu \frac{\theta'(\xi-w)}{\theta(\xi-w)}=0}

It is much more useful to apply slightly shifted definition of the resolvent operator
\be
\boxed{\op(\xi)Z=\left\langle b\sum\lm_i\p_{\xi}\log\theta(\xi-z_i)+\mu\p_{\xi}\log\theta(\xi-w)+2 i a \right\rangle Z}
\ee
Then obviously, the partition function is an eigenfunction of the resolvent integral
\be
\int\lm_{0}^{\pi} d\xi \op(\xi)Z=2\pi i a Z
\ee
Reformulating the loop equation in these terms, one derives
\be
\left[\op^2(z)+Q\p_z(\op(z)-\mu\p_z\log\theta(z-w))-(\zeta(z-w)\p_w-\mu^2\wp(z))+4q\p_q-3\mu(b-\mu)\eta_1\right]Z=0
\ee
where we used the standard elliptic functions \cite{BE}
\be
\zeta(z)=-\p_z\log\theta(z)\nn\\
\wp(z)=\p_z^2\log\theta(z)
\ee

The loop equation possesses a symmetry
\be
\op(\xi)\longrightarrow -\op(\xi)-Q\p_{\xi}\log \op(\xi)-\frac{Q^2}{2}\p_{\xi}\left(\frac{\op'(\xi)}{\op^2(\xi)}\right)-\frac{Q^3}{4}\p_{\xi}\left(-\frac{5}{2}\frac{(\op'(\xi))^2}{\op^4(\xi)}+\frac{\op''(\xi)}{\op^3(\xi)}\right)+O(Q^4)
\ee
Thus, there are two solution branches\footnote{For the toric block all the higher terms are exact so do not contribute 
thus giving the symmetry $\oint\lm_{\gamma}dz\;\op(z)\leftrightarrow -\oint\lm_{\gamma}dz\;\op(z)$.

Notice that for the 4-punctured sphere the second term in the expansion gives a
non-vanishing contribution so the symmetry is $\oint\lm_{\gamma}dz\;\op(z)\leftrightarrow Q-\oint\lm_{\gamma}dz\;\op(z)$ and the symmetric function (invariant) in terms of eigenvalues is the conformal dimension $\Delta(\alpha)=\alpha(Q-\alpha)$, since in this case
\be
\oint_A dz \op(z)Z=\alpha Z
\ee
As usual we switch to a symmetric notation assuming $\alpha=Q/2+a$.}
\be
\oint\lm_{A}dz\;\op^{(+)}(z) Z^{(+)}_{a}=a Z^{(+)}_{a},\quad \oint\lm_{A}dz\;\op^{(-)}(z) Z^{(-)}_{a}=-a Z^{(-)}_{a}
\ee
where the integrals run over the A-period of the spectral surface
and the gauge-invariant quantities like the conformal block may depend only on the invariant 
$\Delta(a)=\frac{Q^2}{4}-a^2$.

\begin{figure}[htbp]
\begin{center}
\includegraphics[scale=0.4]{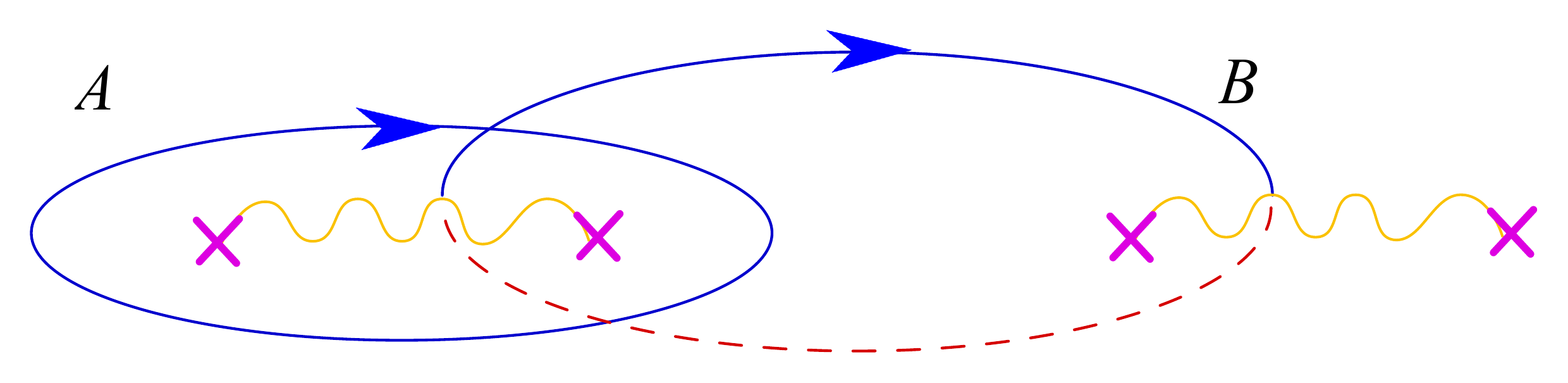}
\caption{Cycles on the spectral curve\label{fig:cycles}}
\end{center}
\end{figure}

\subsection{Resolvents via check-operators \cite{AMMcheck}}

The operator $\hat\nabla(z)$ acting on the matrix model partition function on sphere
inserts $\sum_i \frac{1}{z-z_i}$
inside the correlators (similarly, it inserts $\sum_i \frac{\theta'(\xi-z_i)}{\theta(\xi-z_i)}$ when acting
on the partition function on torus), and therefore is realized as acting
on the infinite set of time variables
in the partition function  entering exponentially the measure, $e^{\sum_{i,k} t_kz_i^k}$:
\be
\hat\nabla(z)
= \sum_k\frac{1}{z^{k+1}}\frac{\partial}{\partial t_k}
\ee
These operators are very convenient in study
of the Ward identities in the form of Virasoro constraints
{\it a la} \cite{AMM}.
However, in the formalism of loop equations,
one usually considers the partition function
with most time variables vanishing, and only a few
left, $t_k=T_k\ne 0$ for $k\le N$ ($N$ is an arbitrary integer parameterizing the class of solutions),
the solution to the Ward identities being parameterized by an arbitrary function of these remaining
variables. Hence, in the formalism of \cite{AMMcheck} the operator $\check\nabla(z)$ can be interpreted
as acting in the moduli space of solutions.
Remarkably, the {\it result} of the action of $\hat\nabla(z)$
(the average of the resolvent) can be represented
as an action of the {\it other} operator
$\check\nabla(z)$, acting only on the moduli space.
This is a somewhat difficult formalism, but it was
developed rather far in \cite{AMMcheck} and we can
now use the results.
For our purposes the main point is that while
\be
\left.\hat\nabla(z|t) Z(t)\right|_{t=T} =
\check \nabla(z|T) Z(T)
\ee
this is not true for repeated action of the resolvent operators:
\be
\left.\hat\nabla(z_1|t)\hat\nabla(z_2|t) Z(t)\right|_{t=T} \neq
\check \nabla(z_1|T)\check \nabla(z_2|T) Z(T)
\ee
Moreover, while $\hat\nabla(z)$ operators at different points $z$
commute,
\be
[\hat\nabla(z_1),\ \hat\nabla(z_2)]\ =\ 0
\ee
this is not true for the check-operators:
\be
[\check\nabla(z_1),\ \check\nabla(z_2)] \ne 0
\ee
The most spectacular result of \cite{AMMcheck} is that
\be\label{PQcheck}
\left[\oint_{A_I} \check \nabla(z),\  \oint_{B^J} \check \nabla(z)\right]
= 2\pi i \delta_I^J
\ee
Actually, \cite{AMMcheck} presented some evidence in favor
of this conjecture by study of the first terms of
the genus expansion only, however, hereafter we assume that
this is true.

To clarify this relation first notice that in the course of calculating the B-cycle integral one has to change the branch.
Thus, it is more safe to consider ``half-cycles" instead.
According to \cite{AMMcheck} the commutation relation for the check-resolvents reads
\be
[\opc(x),\opc(y)]=-\frac{1}{\opc(x)\opc(y)}\left(\p_x-\p_y\right)\frac{\opc(x)^2-\opc(y)^2}{x-y}+{\rm higher\; orders}
\ee

In the case of zero external dimensions, the expectation value of the check-resolvents gives a good spherical approximation
\be
\langle\opc(z)\rangle=\frac{(q(q-1)\p_q\log Z)^{\frac{1}{2}}}{\sqrt{z(z-q)(z-1)}}
\ee
Then, the commutator reads
\be
[\opc(x),\opc(y)]=-\frac{1}{g(x)g(y)}\left(\p_x-\p_y\right)\frac{g(x)^2-g(y)^2}{x-y}+{\rm higher\; orders}
\ee
where $g(z)=\left[z(z-q)(z-1)\right]^{-\frac{1}{2}}$. We introduce notations $A_{1/2}$ and $B_{1/2}$ for the half-cycles
\be
\int\lm_{A_{1/2}}dz=\int\lm_0^q dz,\quad \int\lm_{B_{1/2}}dz=\int\lm_q^1 dz
\ee
Thus the corresponding integral reads
\be
-\int\lm_0^q dx\;\int\lm_q^1\;\frac{1}{g(x)g(y)}\left(\p_x-\p_y\right)\frac{g(x)^2-g(y)^2}{x-y}=\frac{\pi i}{2}
\ee
This simple calculation appeals only to the spherical limit in a simple model, nevertheless, we assume on general grounds
(see also \cite{AMMcheck}) a {\bf non-perturabative} relation
\be
\boxed{\left[\int\lm_{A_{1/2}}dz\;\opc^{(\pm)}(z),\int\lm_{B_{1/2}}dz\;\opc^{(\pm)}(z)\right]=\frac{\pi i}{2}}
\ee

A discussion of the relation between integrals over half-cycles and full cycles we postpone until the consideration of gauge
invariant operators in s. \ref{sec:sur_vs_res}.

Thus, we have constructed the operators manifestly realizing the $pq$-duality. They allow us to construct a dual
pair of the operators $A$ and $B$, the conformal block and the modular transformation like it was done in
section 2.

\subsection{The pair of dual check-operators}

In order to construct the dual pair note that the action of the resolvent (or check-resolvent) operator on the partition
function (conformal block) mimics inserting to the conformal block
the field degenerate at the second level, which can be described in terms of the
$\beta$-ensemble partition function as
\be
Z_{r.op.}(\xi)=\eta^{3\mu (b-\mu)}e^{b\int\lm^\xi du \opc(u)}Z=\eta^{3\mu (b-\mu)}\left\langle\prod\lm_i \theta(\xi-z_i)^{b^2} \theta(\xi-w)^{\mu b}e^{2i a b \xi}\right\rangle Z
\ee
Indeed, the corresponding loop equation can be presented in the form of elliptic Calogero Schr\"odinger equation and coincides
with the two-point conformal block on torus with one field degenerated at the second level, \cite[eq.(34)]{MMM}
\be
\left[4q\partial_q+b^{2}\p_z^2-\left(\zeta(z-w)\p_w+\Delta_{\mu}\wp(z-w)-\Big({3\over 2b^2}+1\Big)\eta_1\right)\right]\tilde Z_{r.op.}(z)=0
\ee
In other words the insertion of an external degenerate field into the conformal block can be, indeed, mimicked literally by the
proper exponential of the check-resolvent.
This gives us an operator expressing the monodromy of a degenerate field along the closed contour $\gamma$:
\be\boxed{
{\cal L}_{\gamma}\sim e^{b\oint\lm_{\gamma} dz\; \opc(z)}\label{eq:sur_fake}}
\ee
It is supposed to represent a ``quantum" version of the abelianization map
discussed in \cite{GMN}.

Now, (\ref{PQcheck}) implies that ${\cal L}_A$ and ${\cal L}_B$ form a pair of dual operators explicitly realized
in $pq$-variables, much similar to the example of s.2. Hence, the corresponding modular transformation
is nothing but the Fourier transform, in accordance with \cite{GMM,Nem}.

\section{Non-perturbative modular transformation}

\subsection{Phase ambiguity}

As we mentioned in our basic example in s.2, to restore the integral kernel, one has to fix the normalization,
which is otherwise is not essential. Therefore, it is important to specify the normalization constant
$N(a)$ relating the partition functions
and the conformal blocks
\be
Z_a(\tau|\mu)=N(a|\mu)B_a(\tau|\mu)
\ee
is essential to determine the modular kernel. Now we consider these normalization constants in details.

\paragraph{Toric normalization constant.}
The normalization constant in the toric case can be simply determined from the partition function integral at $q=0$
\be
N(a)=\int dz_1\ldots \int dz_N\; \prod\lm_{i<j}(\sin z_{ij})^{-2b^2}\prod\lm_i (\sin z_i)^{-2b\mu}\prod\lm_i e^{-4iba z_i}
\ee
After changing the variables $z_i=-\frac{i}{2}\log t_i$, the integral reduces (up to an inessential factor) to
\be
N(a)=\int dt_1\ldots \int dt_N\; \prod\lm_{i<j}t_{ij}^{-2b^2} \prod\lm_i t_i^{-2b \left(a+\frac{Q}{2}\right)}(t_i-1)^{-2b\mu}
\ee
This expression is the Selberg integral which we discuss in Appendix \ref{sec:norm}
\be
N(a)=\tilde N\left(\frac{Q}{2}+a,\mu,\frac{Q}{2}+a\right)=\frac{\left(b^{2b^2-b\mu+1}/\Gamma(-b^2)\right)^{-\frac{\mu}{b}}}{\Gamma_b(0)\Gamma_b(Q-2\mu)}\frac{\Gamma_b(-2a+\mu)\Gamma_b(-2a+Q-\mu)}{\Gamma_b(-2a)\Gamma_b(-2a+Q)}
\ee
where $\Gamma_b (x)$ is the Barnes double gamma function (see Appendix A.1).

\paragraph{4-punctured sphere normalization constant.}
Here we have {\bf an ambiguity} in the definition. Indeed consider {\bf two} $\beta$-ensembles
\be
Z^{(1)}_a=\int\lm_0^x dz_1\ldots\int\lm_0^x dz_{N_1}\boxed{\int\lm_0^1 dz_1\ldots\int\lm_0^1 dz_{N_1}}\;\prod_{i<j}z_{ij}^{-2b^2}\prod_i z_i^{-2b\alpha_0}(z_i-x)^{-2b\alpha_x}(z_i-1)^{-2b\alpha_1}\\
Z^{(2)}_a=\int\lm_0^x dz_1\ldots\int\lm_0^x dz_{N_1}\boxed{\int\lm_1^\infty dz_1\ldots\int\lm_1^\infty dz_{N_1}}\;\prod_{i<j}z_{ij}^{-2b^2}\prod_i z_i^{-2b\alpha_0}(z_i-x)^{-2b\alpha_x}(z_i-1)^{-2b\alpha_1}\nn
\ee
They both give the same expressions for the conformal blocks, though the normalization constants are different
\be
Z^{(1)}_a=\tilde N(\alpha,\alpha_x,\alpha_0)\tilde N(Q-\alpha_\infty,\alpha_1,\alpha)B_a\\
Z^{(2)}_a=\tilde N(\alpha,\alpha_x,\alpha_0)\tilde N(Q-\alpha,\alpha_\infty,\alpha_1)B_a\nn
\ee
Modular kernels are going to be {\bf different} for these two choices. In fact, there is even a larger
ambiguity due to the possibility of using various combinations of the two screening charges (so far we used only one
of them, see \cite{AGTmamoMM})
and there is no \emph{a priori} way to choose between them. One can just say that the modular kernel is defined
up to conjugation with these normalization factors.

\subsection{Non-perturbative dual monodromies from the check-resolvent: toric example}
\label{sec:construction}

In fact, the problem with expression (\ref{eq:sur_fake}) is that it is not gauge-invariant. One could make it
gauge-invariant by taking a trace: a sum of
the both branches
\be
{\cal L}_{\gamma}\sim e^{b\oint\lm_{\gamma} dz\; \opc^{(+)}(z)}+e^{b\oint\lm_{\gamma} dz\; \opc^{(-)}(z)}
\ee
However, the partition functions are different at different branches, thus, one has to switch to the conformal block which is a gauge-invariant object
\be
B_a(\tau|\mu)={Z_a(\tau|\mu)\over N(a)}
\ee
This means one has to twist the exponentials of the check operator by the corresponding normalization constants
\be
e^{b \oint dz\; \opc(z)}\longrightarrow {1\over N(a)} e^{b \oint dz\; \opc(z)} N(a)
\ee
The branches differ by the sign of the check operator, thus, ultimately the relation between the monodromy operator
and the exponential of the check-resolvent reads
\be\label{50}
\boxed{{\cal L}_{\gamma}^{\rm tor}=\left[\frac{1}{N(a)}e^{b\oint\lm_{\gamma}dz\;\opc(z)}N(a)+
\frac{1}{N(-a)}e^{-b\oint\lm_{\gamma}dz\;\opc(z)}N(-a)\right]}
\ee
This operator is well-defined on the whole moduli space and the conformal blocks are eigenvectors of its A-periods.

We need only the $a$-dependent part of the normalization constant:
\be
N(a)=\frac{\Gamma_b(2a+\mu)\Gamma_b(2a+Q-\mu)}{\Gamma_b(2a)\Gamma_b(2a+Q)}
\ee
Substituting the $a$-representation for the check operator,
\be
\oint\lm_A dz\; \opc(z)=2\pi i a,\quad \oint\lm_B dz\; \opc(z)=\frac{1}{2} \p_a
\ee
one derives
\be\label{donp}
{\cal L}_A=2\cos 2\pi b a\\
{\cal L}_B=\frac{\Gamma(2 a b)\Gamma(b Q+2 a b)}{\Gamma( b\mu +2  a b)\Gamma(b(Q-\mu) +2  a b)}e^{\frac{b}{2}\p_a}+\frac{\Gamma(-2 a b)\Gamma(b Q-2 a b)}{\Gamma( b\mu -2  a b)\Gamma(b(Q-\mu) -2  a b)}e^{-\frac{b}{2}\p_a}
\nn
\ee

Thus, we have constructed the two operators, ${\cal L}_A$ and ${\cal L}_B$ from the check operators with
the canonical commutation relations. They provide the exchange relation in the CFT, that is, the modular transformation.
Hence, the modular invariance is {\bf a transformation induced by the $pq$-duality}.
Moreover, in the perturbative regime (i.e. at large $a$)
these operators contain only one of the two exponentials associated with one of the two branches, i.e. the modular
transformation in this regime, indeed, reduces to the Fourier transformation \cite{GMM,Nem} as we discussed in the
previous section (the pre-exponential factor in (\ref{50}) in this case, when only one of the exponentials survives
is absorbed into the normalization of the conformal block).

\subsection{Check and surface operators}
\label{sec:surface}

The dual operators ${\cal L}_{A,B}$ possess also an interpretation as surface operators \cite{AGGTV,DGOT}. The explicit
expressions for them were already obtained in \cite{AGGTV,DGOT} by some heuristic arguments, and they coincide with the
result of our straightforward calculation in the previous subsection.

More concretely, the two fields degenerate at the second level of the Virasoro algebra have the following OPE:
\be
\Phi_{(2,1)}\otimes\Phi_{(2,1)}=\Phi_{(1,1)}\oplus \Phi_{(3,1)}
\ee
And the field $\Phi_{(1,1)}$ has dimension 0 and can be thought as an operator acting in the space of conformal blocks.
In other words one can perform the following operation $\cal C$ mapping $n$-point conformal blocks $CB_n$ to the
degenerate $n+2$-point blocks constructing a solution to the equation
\be
{\cal C}: CB_{n}\longrightarrow CB_{n|2}\\
(b^2L_{-1}^2-L_{-2})\left\langle V_{b/2}(z)V_{b/2}(w){\cal O}\right\rangle=0,\quad \left\langle V_{b/2}(z)V_{b/2}(w){\cal O}\right\rangle\sim (z-w)^{\frac{b^2}{2}}\left\langle {\cal O}\right\rangle
\ee
Using the same differential equation, one can generate a monodromy transformation making a parallel transport of one of the degenerate fields along some contour $\gamma$:
\be
{\cal M}_{\gamma}: CB_{n|2}\longrightarrow CB_{n|2}
\ee
In this way, one constructs the Verlinde (surface) \cite{AGGTV,DGOT,Verlinde,surface} operator
\be
{\cal L}_{\gamma}={\cal C}^{-1}{\cal M}_{\gamma}{\cal C}: CB_n \longrightarrow CB_n
\ee
It is important to show that this operator can be formulated as a differential operator acting on the conformal block,
at least in some abstract form. This makes the Verlinde operator quite similar
to the check operator constructed within the matrix model framework and means that the
Verlinde operator is a kind of exponential of the check
operator. This should be compared with what we did in \cite{MMM} considering a slightly different operator
${\cal L}_{\gamma}=(C_x')^{-1}{\cal M}_{\gamma} C_x'$
\be
C_x': CB_n \longrightarrow CB_{n|1}
\ee
and constructing it as a solution to the following equation
\be
\left[b^2z(z-1)\p_z^2-(2z-1)\p_z-\frac{x(x-1)}{z-x}\p_x+\Delta_{1/2b}+\frac{\Delta_0}{z}-\frac{\Delta_1}{z-1}-\Delta_\infty +
\frac{x^2-(2x-1)z}{(z-x)^2}\Delta_x \right]B_{4|1}(z|x)=0,\nn\\
B_{4|1}(z|x)=B_4(x)(z-x)^{\frac{Q}{2}-\sqrt{\frac{Q^2}{4}-\Delta}}\left(1+O\left((z-x)\right)\right)\hspace{5cm}
\ee
This is the equation for the 5-point conformal block with one field degenerate at the second level and with the corresponding intermediate dimension fixed,
see \cite[eqs.(26,30)]{MMM} for details.

An explicit expression for those operators can be found in terms of CFT \cite{AGGTV,DGOT}, not only for the
one-point toric but also for the four-point spherical conformal blocks:
\begin{itemize}
  \item Toric:
\be \label{eq:CFT_tor}
{\cal L}_A=2\cos 2\pi b a\\\nn\\
{\cal L}_B=\frac{\Gamma(2 a b)\Gamma(b Q+2 a b)}{\Gamma( b\mu +2  a b)\Gamma(b(Q-\mu) +2  a b)}e^{\frac{b}{2}\p_a}+\frac{\Gamma(-2 a b)\Gamma(b Q-2 a b)}{\Gamma( b\mu -2  a b)\Gamma(b(Q-\mu) -2  a b)}e^{-\frac{b}{2}\p_a}
\nn
\ee
  \item 4-punctured sphere:
    \be
    {\cal L}_A=\cos 2\pi b a,\\
    {\cal L}_B=H_+(a)e^{b\p_a}+H_0(a)+H_-(a)e^{-b\p_a},\nn
    \ee
    where
    \be\label{eq:CFT_pm}
    H_\pm(a)=4\pi^2\frac{\Gamma\left(b(Q/2\pm 2a+b)\right)\Gamma\left(b(Q/2\pm 2a)\Gamma\left(b(\pm 2a+b)\right)\Gamma\left(b(\pm 2a)\right)\right)}{\prod\lm_{s_i=\pm}\Gamma\left(b(Q/2\pm a+s_1\mu_1+
    s_2\mu_2)\right)\Gamma\left(b(Q/2\pm a+s_3\mu_3+s_4\mu_4)\right)}\\
    \begin{split}\label{eq:CFT_0}
    H_0(a)=\frac{\cos\pi b^2}{\cos 4\pi b a-\cos 2\pi b^2 }(\cos 2\pi b \mu_2\cos 2\pi b \mu_3+\cos 2\pi b \mu_1 \cos 2\pi b \mu_4)+\\
    +\frac{\cos 2\pi b a}{\cos 4\pi b a-\cos 2\pi b^2 }(\cos 2\pi b \mu_1\cos 2\pi b \mu_3+\cos 2\pi b \mu_2 \cos 2\pi b \mu_4)
    \end{split}
    \ee
\end{itemize}
and the variables $\mu_i$ are related to the conformal dimensions in the 4-point spherical conformal block case as
$\Delta_i=\mu_i(Q-\mu_i)$. Notice that our normalization for $H_{\pm}$ differs from \cite{AGGTV} by $2\pi$.

The result for the toric case coincides with formula (\ref{donp}) obtained in the previous subsection.

\subsection{Towards the four-punctured sphere example}\label{sec:sur_vs_res}

In the case of a punctured sphere our approach of s.5.2 becomes more subtle. The problem is that one has
to switch branches while going along the B-cycle (see Fig.\ref{fig:cycles}). However, having constructed
the gauge-invariant operator, we expect a natural relation ${\cal L}_{\gamma}\sim
{\cal L}_{\gamma_{1/2}}{\cal L}_{\gamma_{1/2}}$,
where $\gamma_{1/2}$ denotes a ``half" of the contour going just along one branch (either solid or dashed line on
Fig.\ref{fig:cycles}), though one can not exclude appearance of trace terms in this expression. Hence,
generally this operator expansion reads
\be
{\cal L}_{\gamma}=c_1 {\cal L}_{\gamma_{1/2}}{\cal L}_{\gamma_{1/2}}+c_2
\ee
The unknown coefficients $c_1$ and $c_2$ can be easily read off from the relation for monodromies along the A-cycle
(\ref{donp}).
Indeed, both the operator and the ``half-operator" are well-defined
\be
{\cal L}_A=2\cos 2\pi b a, \quad {\cal L}_{(0,x)}=2\cos \pi b a
\ee
Implementing a simple trigonometric identity $\cos 2x=2\cos^2 x-1$, one states the realization of the monodromy 
operator as a check operator
\be
\boxed{{\cal L}_{\gamma}^{\rm 4-pun}=\left[\frac{1}{N(a)}e^{b\int\lm_{\gamma_{1/2}}dz\;\opc(z)}N(a)+\frac{1}{N(-a)}e^{-b\int\lm_{\gamma_{1/2}}dz\;\opc(z)}N(-a)\right]^2-2}
\ee
This expression allows one to compute the shifting coefficients explicitly
\be
{\cal L}_{\gamma}=H_+(a)e^{b\p_a}+H_0(a)+H_-(a)e^{-b\p_a},\\
H_{\pm}(a)=\frac{N(\pm a+b)}{N(\pm a)},\quad H_0(a)=\frac{N( a+b/2)}{N( a)}\frac{N( -a)}{N( -a-b/2)}+\frac{N(- a+b/2)}{N(- a)}\frac{N( a)}{N( a-b/2)}-2
\ee
Now one suffices to substitute explicit expressions for $N(a)$ in order to obtain the final answer.
However, as we already emphasized in s.5.1, there is an ambiguity in the normalization factor in this case. Hence, at the
moment we just read off $N(a)$ from the known surface operator (\ref{eq:CFT_pm}), leaving a discussion of
this subtle point for a separate publication.

If one chooses
\be\label{eq:norm}
N(a)=\frac{\prod\lm_{s_i=\pm}\Gamma_b(Q/2+a+s_1\mu_1+s_2\mu_2)\Gamma_b(Q/2+a+s_3\mu_3+s_4\mu_4)}{\Gamma_b(2a+Q)\Gamma_b(2a)}
\ee
this gives the value of $H_\pm$ coinciding with (\ref{eq:CFT_pm}). Then, the real challenge is to reproduce the
magnetic term contribution $H_0$. Formula (\ref{eq:norm}) can not be appropriate for this purpose, since it is
symmetric under permutation of $\mu_1$ and $\mu_2$, and formula (\ref{eq:CFT_0}) is not.  Note, however, that
(\ref{eq:norm}) can be multiplied by any periodic function of $a$ with period $b$, which does not effect $H_\pm$,
while changing $H_0$.

Note also that at particular values $\mu_1=\mu_4=\frac{b}{4}$, the correct answer for $H_0$ is obtained directly
from (\ref{eq:norm}). Indeed, in this case
\be
\frac{\Gamma_b(x+\mu_1+b/2)\Gamma_b(x-\mu_1+b/2)}{\Gamma_b(x+\mu_1)\Gamma_b(x-\mu_1)}=
\frac{\sqrt{2\pi}b^{b(x-b/4-1/2)}}{\Gamma(bx-b^2/4)}
\ee
Applying this relation one finds the ratio
\be
F(a)=\frac{N( a+b/2)}{N( a)}\frac{N( -a)}{N( -a-b/2)}=-4\frac{ \prod _{s_i=\pm} \cos \left(\pi  b \left(a+\frac{b}{4}+s_2 \mu_2\right)\right)\cos \left(\pi  b \left(a+\frac{b}{4}+s_3 \mu_3\right)\right)}{\sin (2 \pi  a b) \sin (\pi  b (2 a+b))}
\ee
and, after a simple algebra, one indeed obtains
\be
F(a)+F(-a)-2=H_0\left(a,\mu_1=\frac{b}{4},\mu_2,\mu_3,\mu_4=\frac{b}{4}\right)
\ee

\section{Modular kernel non-perturbatively}
\label{sec:modular}

In this section we demonstrate that the modular kernel can be straightforwardly read off from the equation
\be\label{ref}
{\cal L}_B(a)M(a,a')={\cal L}_A(a') M(a,a')
\ee
much similar
to eq.(\ref{kernel}) of section 2. Let us consider the toric one-point conformal block, i.e. formulas
(\ref{donp}). Note that the source of complexity of the modular kernel is a complicated structure of the conformal
block  asymptotic series $N(a)\neq N(-a)$. Let us be more specific in this place: divide the normalization factor
in symmetric and non-symmetric parts $N(a)=N_n(a)N_s(a)$, where $N_s(-a)=N_s(a)$. Then, the monodromy
operators can be simplified
\be
{\cal L}_{\gamma}={1\over N_n(a)N_s(a)} e^{b\oint\lm_{\gamma}dz\opc(z)}N_n(a)N_s(a)+{1\over
N_n(-a)N_s(-a)} e^{-b\oint\lm_{\gamma}dz\opc(z)}N_n(-a)N_s(-a)=N_s(a)^{-1}{\cal L}_{\gamma}'N_s(a),\nn\\
{\cal L}_{\gamma}'=N_n(a)^{-1} e^{b\oint\lm_{\gamma}dz\opc(z)}N_n(a)+N_n(-a)^{-1} e^{-b\oint\lm_{\gamma}dz\opc(z)}N_n(-a)
\ee
since $N_s(a)$ being Weyl symmetric coincides on different branches. Let us split the normalization factor
\be
\begin{split}
N(a)=\frac{\Gamma_b(2a+\mu)\Gamma_b(2a+Q-\mu)}{\Gamma_b(2a)\Gamma_b(2a+Q)}=\frac{\Gamma_b(2a+\mu)\Gamma_b(Q-2a)}{\Gamma_b(2a)\Gamma_b(Q-2a-\mu)}\frac{\Gamma_b(2a+Q-\mu)\Gamma_b(-2a+Q-\mu)}{\Gamma_b(2a+Q)\Gamma_b(-2a+Q)}=\\
=\frac{S_b(2a+\mu)}{S_b(2a)}\left[\frac{\Gamma_b(2a+Q-\mu)\Gamma_b(-2a+Q-\mu)}{\Gamma_b(2a+Q)\Gamma_b(-2a+Q)}\right]
\end{split}
\ee
where $S_b(x)$ is the double sine function (see Appendix A.2), and
we throw away the last symmetric multiplier in the brackets\footnote{In fact, the expressions for the surface
operators in \cite{DGOT} and \cite{AGGTV} differ exactly by this symmetric factor.}. Then, we obtain (cf. with
\cite[eq.(5.25)]{DGOT})
\be\label{newdo}
N_a(a)=\frac{S_b(2a+\mu)}{S_b(2a)},\\
{\cal L}_A'=\cos 2\pi b a,\quad {\cal L}_B'=\frac{1}{2}\left(\frac{\sin 2\pi b(a-\mu/2)}{\sin 2\pi b a}e^{-\frac{1}{2}b\p_a}+
\frac{\sin 2\pi b(a+\mu/2)}{\sin 2\pi b a}e^{\frac{1}{2}b \p_a}\right)
\ee
Now one can solve the eigenvalue problem using expressions (\ref{newdo}) in (\ref{ref}):
%For the sake of simplicity, we consider the particular case of $b=1$ (i.e. $c=25$)
\be
\frac{1}{2}\left(\frac{\sin 2\pi b(a-\mu/2)}{\sin 2\pi ba}e^{-\frac{b}{2}\p_a}+
\frac{\sin 2\pi b(a+\mu/2)}{\sin 2\pi ba}e^{\frac{b}{2}\p_a}\right)M(a,a')=\cos 2\pi ba'\; M(a,a')
\ee
It is simpler to solve this equation after performing the Fourier transform
\be
M(a,a')=\int\lm_{-\infty}^{\infty}d\xi e^{4\pi i a\xi}f_{a'}(\xi)
\ee
This leads to the substitution
\be
e^{2\pi i ba}\longrightarrow e^{-\frac{b}{2}\p_{\xi}},\quad e^{\frac{b}{2}\p_a}\longrightarrow e^{2\pi i b\xi}
\ee
and we use the following variables
\be
\eta=e^{\pi i b^2},\quad y=e^{\pi i b\mu},\quad z=e^{2\pi i b\xi},\quad s=e^{2\pi i ba'},\quad \hat X f(\xi)=e^{\frac{b}{2}\p_\xi}f(\xi)
\ee
Then, the eigenvalue problem reduces to the following algebraic equation
\be
\left[\Big(\hat X y-\hat X^{-1}y^{-1}\Big)z^{-1}+\Big(\hat X y^{-1}-\hat X^{-1}y\Big)z\right]f_{a'}(\xi)=
\Big(\hat X-\hat X^{-1}\Big)\Big(s+\frac{1}{s}\Big)f_{a'}(\xi) \Rightarrow\nn
\\
\Rightarrow\quad \hat X^2f_{a'}(\xi)=\eta^2\frac{(s-yz)(1-zys)}{(sy-\eta^2z)(y-\eta^2sz)}f_{a'}(\xi)
\ee
or, equivalently, to
\be
f_{a'}(\xi+b)=\frac{\sin\pi b\left(\xi+\frac{\mu}{2}-a'\right)\sin\pi b\left(\xi+\frac{\mu}{2}+
a'\right)}{\sin\pi b\left(\xi+b-\frac{\mu}{2}-a'\right)\sin\pi b\left(\xi+b-\frac{\mu}{2}+a'\right)}f_{a'}(\xi)
\ee
The solution reads
\be
f_{a'}(\xi)=\tilde C_1(\xi)C_2(a')\frac{S_b\left(\xi+\frac{\mu}{2}-a'\right)S_b\left(\xi+\frac{\mu}{2}+a'\right)}{S_b\left(\xi
+b-\frac{\mu}{2}-a'\right)S_b\left(\xi+b-\frac{\mu}{2}+a'\right)}=C_1(\xi)C_2(a')\frac{S_b\left(\xi+\frac{\mu}{2}-a'\right)
S_b\left(\xi+\frac{\mu}{2}+a'\right)}{S_b\left(\xi
+Q-\frac{\mu}{2}-a'\right)S_b\left(\xi+Q-\frac{\mu}{2}+a'\right)}\nn
\ee
where $C_1(\xi)$ is an 
arbitrary periodic function with period $b$, $C_2(a')$ is an arbitrary function 
and we used the fact that the function $G(x)=
\displaystyle{e^{\pi ix/b}{S_b(x+1/b)\over
S_b(x)}}$ is periodic with period $b$. Thus, finally,
\be
\boxed{M(a,a')=\int d\xi\; C_1(\xi)C_2(a')\frac{S_b\left(\xi+\frac{\mu}{2}-a'\right)S_b\left(\xi+\frac{\mu}{2}+
a'\right)}{S_b\left(\xi+Q-\frac{\mu}{2}-a'\right)S_b\left(\xi+Q-\frac{\mu}{2}+a'\right)}e^{4\pi i a\xi}}%=\\
%\nn\\
%=e^{-2\pi i a Q}\int d\xi\; C_1(\xi)C_2(a')\frac{S_b\left(\xi+\frac{\mu-Q}{2}-a'\right)S_b\left(\xi+\frac{\mu-Q}{2}+
%a'\right)}{S_b\left(\xi-\frac{\mu-Q}{2}-a'\right)S_b\left(\xi-\frac{\mu-Q}{2}+a'\right)}e^{4\pi i a\xi}
\nn
\ee
and there is a freedom in this answer related with the choice of normalization of the conformal block. 
This result is consistent\footnote{In order to compare the two answers, one has to use property (\ref{dss}).}
with \cite[(4.41)]{Tesch}, and here it is obtained by solving directly the simple and explicit equation
(\ref{ref}). 

\section*{Acknowledgements}

We are grateful for the hospitality at the International Institute of Physics, UFRN (Brazil),
where the paper was completed. 
Our work is partly supported by the DOE under grant DE-FG02-96ER40959 (D.G.), by
Ministry of Education and Science of the Russian Federation, the Brazil National Counsel of Scientific and
Technological Development (A.Mor.), by the program  of UFRN-MCTI, Brazil (A.Mir.),
by NSh-3349.2012.2, by RFBR grants 13-02-00457 (D.G. and A.Mir.), 13-02-00478 (A.Mor.), by joint grants 12-02-92108-Yaf,
13-02-91371-ST, 14-01-93004-Viet, 14-02-92009-NNS and by leading young scientific groups RFBR 12-02-315353-mol-a (D.G.).

\appendix

\section{Useful quantum functions}

\label{sec:dilog}
In this Appendix we list some useful definitions of the Barnes functions. We follow conventions of \cite{DGOT}.

\subsection{The double gamma function $\Gamma_b(x)$}

This function satisfies the functional equation
\be
\Gamma_b(x+b)={\sqrt{2\pi}b^{bx-\frac{1}{2}}\over\Gamma(bx)}\Gamma_b(x)
\ee
with the ordinary $\Gamma$-function in the denominator, and possesses the integral representation
\be
\log \Gamma_b(x)=\int\lm_{0}^{\infty}\frac{dt}{t}\; \left(\frac{e^{-xt}-e^{-Qt/2}}{\left(1-e^{bt}\right)\left(1-e^{t/b}\right)}-\frac{(Q-2x)^2}{8e^t}-\frac{Q-2x}{t}\right)
\ee
which immediately implies
\be
\Gamma_b(x)=\Gamma_{1/b}(x)
\ee

\subsection{The double sine function $S_b(x)$}

This function is defined as
\be
S_b(x)=\frac{\Gamma_b(x)}{\Gamma_b(Q-x)}
\ee
satisfies the difference equation
\be
S_b(x+b)=2\sin \pi b x\; S_b(x)
\ee
and enjoys the evident property
\be\label{dss}
S_b(x)S_b(Q-x)=1
\ee

\section{Normalization of the matrix model partition function}
\label{sec:norm}
The ``holomorphic" three-point correlation function $\tilde N$ is defined through the Selberg integral
\be
\tilde N(\alpha_3,\alpha_2,\alpha_1)=\frac{1}{(2\pi)^N N!}\prod\lm_{i=1}^N \int\lm_0^1 dz_i\; \prod\lm_{i<j}z_{ij}^{-2b^2}\prod\lm_{i=1}^N z_i^{-2b\alpha_1}(1-z_i)^{-2b \alpha_2},\nn\\
\alpha_1+\alpha_2+bN=\alpha_3
\ee
For any integer $N$ this integral can be calculated explicitly
\be
\tilde N(\alpha_3,\alpha_2,\alpha_1)=\prod\lm_{j=1}^N\frac{\Gamma\left(-2b\alpha_1+1-b^2(j-1)\right)\Gamma\left(-2b\alpha_2+1-b^2(j-1)\right)\Gamma(1-b^2 j)}{\Gamma\left(-2b\alpha_1-2b\alpha_2+2-b^2(N+j-2)\right)\Gamma(1-b^2)}
\ee
Using the functional relation
\be
\Gamma(bx)=\sqrt{2\pi}b^{bx-\frac{1}{2}}\frac{\Gamma_b(x)}{\Gamma_b(x+b)}
\ee
one can derive its analytic continuation to arbitrary $N$ \cite{three-point}
\be
\tilde N(\alpha_1,\alpha_2,\alpha_3)=\left(\frac{b^{(N+2)b^2+1}}{\Gamma(-b^2)}\right)^N\frac{\Gamma_b (2 Q-\alpha_1-\alpha_2-\alpha_3) \Gamma_b (Q-\alpha_1+\alpha_2-\alpha_3) \Gamma_b (Q-\alpha_1-\alpha_2+\alpha_3) \Gamma_b (-\alpha_1+\alpha_2+\alpha_3)}{\Gamma_b (2 Q-2\alpha_1) \Gamma_b (Q-2\alpha_2) \Gamma_b (Q-2\alpha_3)\Gamma_b(0)}
\nn
\ee

\end{document}